\documentclass{aastex631}

\newcommand{\sbt}{\,\begin{picture}(-1,1)(-1,-3)\circle*{3}\end{picture}\ }
\usepackage{array}

\begin{document}
\title{The Influence of Different Phases of a Solar Flare on Changes in the Total Electron Content in the Earth's Ionosphere}

\correspondingauthor{Susanna Bekker}
\email{s.bekker@qub.ac.uk, susanna.bekker@gmail.com}

\author[0000-0001-5302-2543]{Susanna Bekker}
\affiliation{Astrophysics Research Centre, School of Mathematics and Physics, Queen’s University Belfast \\ University Road \\ Belfast, BT7 1NN, UK}

\author[0000-0001-5031-1892]{Ryan O. Milligan}
\affiliation{Astrophysics Research Centre, School of Mathematics and Physics, Queen’s University Belfast \\ University Road \\ Belfast, BT7 1NN, UK}

\author[0000-0003-4975-4164]{Ilya A. Ryakhovsky}
\affiliation{Sadovsky Institute of Geospheres Dynamics, Russian Academy of Sciences \\ 38 build 1 Leninsky avenue \\ Moscow, 119334, Russia}

\begin{abstract}

Variations in $X$-ray and EUV irradiance during solar flares lead to a noticeable increase in the electron concentration in the illuminated part of the Earth's ionosphere. Due to the large amount of experimental data accumulated by Global Navigation Satellite Systems (GNSS), the total electron content ($TEC$) response to the impulsive phase of a solar flare has been studied quite well. However, recent studies have shown that large fraction of $X$-class flares have second strong peak of warm coronal emission (which is called “EUV late phase”), whose influence on the ionization of ionospheric layers is not yet clear. A combined analysis of successive solar emissions and the caused $TEC$ changes made it possible to numerically estimate the ionospheric response to the impulsive, gradual, and late phases of the $X$2.9 solar flare occurred on 2011 November 3 and demonstrate the high geoeffectiveness of the rather weak \ion{Fe}{15} 28.4 nm solar emission during the EUV late phase. It was found that the ionospheric response to the relatively weak emissions of the EUV late phase of the $X$2.9 solar flare amounted to almost a third of the $TEC$ increase during the impulsive phase.

\end{abstract}


\section{Introduction} 

The first record of the impact of solar flares on the Earth’s ionosphere was made a century and a half ago \citep{Curto_2020}. The fast development of ground-based and space-based technologies in the 20$^{th}$ century led to the accumulation of a significant amount of experimental information on the behavior of ionospheric layers under the influence of solar radiation of various power, spectral composition, duration, and periodicity \citep{Meza_etal_2006,Guyer_Can_2013,Hazarika_etal_2016,Barta_etal_2019,Qian_etal_2019,Habarulema_etal_2022,Buzas_etal_2023,Nishimoto_2023}. The discovered empirical patterns resulted in a significant progress in understanding the physics of solar-terrestrial connections and enabled to predict the possible response of the Earth's ionosphere to sudden changes in solar radiation. However, new information has also given rise to new questions about the ionization of the medium under quiet and disturbed conditions, for example, during a sharp increase in ionization caused by solar flares of various power. Therefore, modeling and interpreting geophysical responses to variations in solar plasma parameters is still a relevant problem which also has an important impact on modern communication and navigation systems.

A solar flare is a sudden event on the Sun that releases enormous amounts of energy. Variations in solar radiation in different emission lines and continua caused by solar flares lead to a non-uniform increase in the electron concentration ($Ne$) at different ionospheric altitudes \citep{Mitra_1974,Leonovich_etal_2002}. $X$-ray emission ($\lambda <$ 10 nm) generally comes from hot plasma ($\sim10^7$ K) confined to coronal loops. It is known that during flares the $X$-ray flux increases by orders of magnitude \citep{Basak_Chakrabarti_2019}, penetrates into the lower ionosphere layers and becomes the main source of $D$ region ionization ($h <$ 90 km). This leads to perturbation of the characteristics of VLF signals (3–30 kHz), which propagate in the waveguide between the Earth's surface and the ionospheric $D$ region \citep{Thomson_Clilverd_2001,Palit_etal_2013,Raulin_etal_2013,Singh_etal_2014,Hayes_etal_2021,Nina_2022,Bekker_Korsunskaya_2023}. The amplitude and phase of VLF signals are very sensitive to variations in $Ne$, and therefore are widely used to study the dynamics of the lower ionosphere under various heliogeophysical conditions, including for constructing empirical models of the $D$ region \citep{Wait_Spies_1974,Thomson_1993,Ferguson_1998,Gavrilov_etal_2019a,Nina_2022} and verifying theoretical ones \citep{Chowdhury_etal_2021,Bekker_etal_2022}. Extreme ultraviolet (EUV) emission comes predominantly from the loop footpoints formed in the solar chromosphere ($\sim10^4$$-$$10^5$ K). It has more moderate fluctuations but is absorbed at altitudes of more ionized $E$ (90 $< h <$ 120 km) and $F$ ($h >$ 120 km) regions of the ionosphere, where the maximum electron density is located. Therefore, it is mostly the EUV range that is responsible for the increase in total electron content ($TEC$) in the ionosphere during flares \citep{Hazarika_etal_2016,Leonovich_etal_2002,Tsurutani_etal_2009}.

The spectral coverage of extreme ultraviolet observations obtained by the SDO satellite (Solar Dynamics Observatory; \citealp{Pesnell_etal_2012}) has enabled extensive analysis of the different phases of a solar flare \citep{Woods_etal_2011}. As the $X$-ray emitting plasma cools during the gradual phase of the flare, there can also be successive EUV emission at intermediate temperatures ($\sim10^5$$-$$10^6$ K). \citet{Woods_etal_2011} showed that one variant of coronal loop reconnection after the impulsive phase of a flare results in warm coronal emissions (such as \ion{Fe}{16} 33.5 nm, $T$ = $10^{6.43}$ K) having a second large emission peak that can lag the primary flare by hours. This peak is called the EUV late phase \citep{Woods_etal_2011,Liu_etal_2015,Dai_etal_2018,Zhou_etal_2019,Chen_etal_2020,Zhong_etal_2021,Liu_etal_2024,Liu_etal_2024_2}. The influence of the described successive $X$-ray and ultraviolet emissions on the dynamics of the Earth's ionosphere parameters has not yet been fully studied and is very important for understanding the complex nature of the ionospheric response to variations in solar radiation.

One of the most valuable tools for studying solar-terrestrial connections is Global Navigation Satellite Systems (GNSS). Long-term studies devoted to the experimental assessment of $TEC$ (using data on the phase and code delay of GPS signals) made it possible to investigate the latitude-longitudinal variations in electron density depending on the season, time of day and other heliogeophysical conditions \citep{Tsurutani_etal_2005,Qian_etal_2019,Alizadeh_2020,Watanabe_etal_2021}. An increase in electron density during solar flares leads to a delay in GPS signals, so GNSS data can be used to quantify changes in the total electron content in the ionosphere \citep{Wan_etal_2002,Wan_etal_2005,Garcia_etal_2007,Le_etal_2013,Yasyukevich_etal_2018,Habarulema_etal_2020}. \citet{Meza_etal_2006,Hazarika_etal_2016,Todorovi_etal_2016} and \citet{Gavrilov_etal_2019b} showed that variations in the total electron content during powerful solar flares amount to several tecu (1 tecu $= 10^{16} m^{-2}$), and at low latitudes they can even exceed 10 tecu \citep{Yasyukevich_etal_2018}, which is approximately 50\% of the average global total electron content during the low solar activity \citep{Lean_etal_2016}. Such strong fluctuations often cause problems with radio communications and, in particular, strong absorption of HF radio waves \citep{Yasyukevich_etal_2018,Habarulema_etal_2020}.

It is worth noting that the value of the total electron content is integral and does not provide detailed information about the change in the vertical profile of $Ne$ caused by natural disturbances. At the same time, the temporal dynamics of the total electron content increment contains information about the electron concentration response to various emissions of the solar spectrum during the impulsive, gradual, and late phases of a solar flare. Moreover, it can be an experimental source of information about the geoeffectiveness of various lines and continua of the solar spectrum.

Most modern works have been devoted to estimating $TEC$ changes during the impulsive phase of a solar flare. There is only one published work that presents estimates of the total electron content response to the EUV late phase. \citet{Liu_etal_2024} numerically compared $TEC$ values during a quiet pre-flare day (2012 October 22) with $TEC$ values during an $X$-class flare, which was characterized by a pronounced late phase in the \ion{Fe}{16} line (2012 October 23), and found a maximum difference of $\sim5$ tecu near the subsolar point. This value decreased with increasing distance from the subsolar point. The obtained result demonstrated the high relevance of studying the ionospheric reaction not only to the impulsive phase of the flare, but also to later emissions.

This paper proposes another approach to assessing the ionospheric response to the EUV late phase on the example of the $X2.9$ solar flare occurred on 2011 November 3. Here the estimation of the $TEC$ response to various phases of a solar flare was calculated under next conditions:

\begin{itemize}
  \item[\sbt] to obtain a clear $Ne$ increase caused only by the EUV late phase, the $\Delta TEC$ is assessed by analyzing the dynamics of slant $TEC$ along the path from the satellite to the receiver, and not variations in absolute values;
  \item[\sbt] due to the chosen approach, the dynamics of $\Delta TEC$ is considered with a high cadence (15 seconds) in order to track the $Ne$ response to specific solar radiation emissions.
\end{itemize}

The purpose of this study is to experimentally estimate the response of the total electron content to various phases of an $X$-class solar flare and analyze the temporal structure of $TEC$ variations to determine the most geoeffective emissions from the solar spectrum. To avoid the effect of random phenomena that are not related to solar radiation emissions at various flare phases, a statistical analysis of the results obtained at many GNSS stations is carried out.

In Section 2, the solar data used and the spectrum of the specific flare are described. Section 3 presents ionospheric data and a methodology for estimating total electron content increment. The results obtained and their discussion are presented in Section 4 of this paper.

\section{Solar data}

\subsection{SDO/EVE and SDO/AIA Flare Observations}

Solar EUV spectral irradiance measurements were obtained using the EVE (Extreme ultraviolet Variability Experiment; \citealp{Woods_etal_2012}) instrument located on the SDO satellite. Since 2010, it has been monitoring variations in solar radiation with high temporal (10 seconds) and moderate spectral resolution (0.1 nm). The data obtained enable a detailed analysis of the emissions during solar flares of various classes and spectral composition. Unfortunately, spectral measurements of the entire wavelength range 0.1–106 nm obtained by both the SDO/EVE MEGS (Multiple EUV Grating Spectrographs) instruments, MEGS-A and MEGS-B, are available only until May 2014. However, this period also included a sufficient number of flares of various classes (including 45 $X$-class flares) that can be used for combined analysis of solar emissions and the corresponding ionospheric response. Some of these events had a pronounced EUV late phase, which, according to the definition proposed by \citet{Woods_etal_2011}, must satisfy the following conditions:
\begin{itemize}
  \item[\sbt] this should be a second peak of the warm coronal emissions (\ion{Fe}{15} and \ion{Fe}{16}), which occurs from several minutes to several hours after the $X$-ray peak;
  \item[\sbt] there is no significant enhancement of $X$-rays or hot coronal emissions (such as \ion{Fe}{10}/\ion{Fe}{13} 13.3 nm) during a late phase;
  \item[\sbt] an eruptive event (as coronal dimming in the \ion{Fe}{9} 17.1 nm) should be observed;
  \item[\sbt] the second set of post-eruptive coronal loops should be higher than the first set of post-flare loops.
\end{itemize}

\begin{figure} [h]
        \noindent\includegraphics[width=470pt]{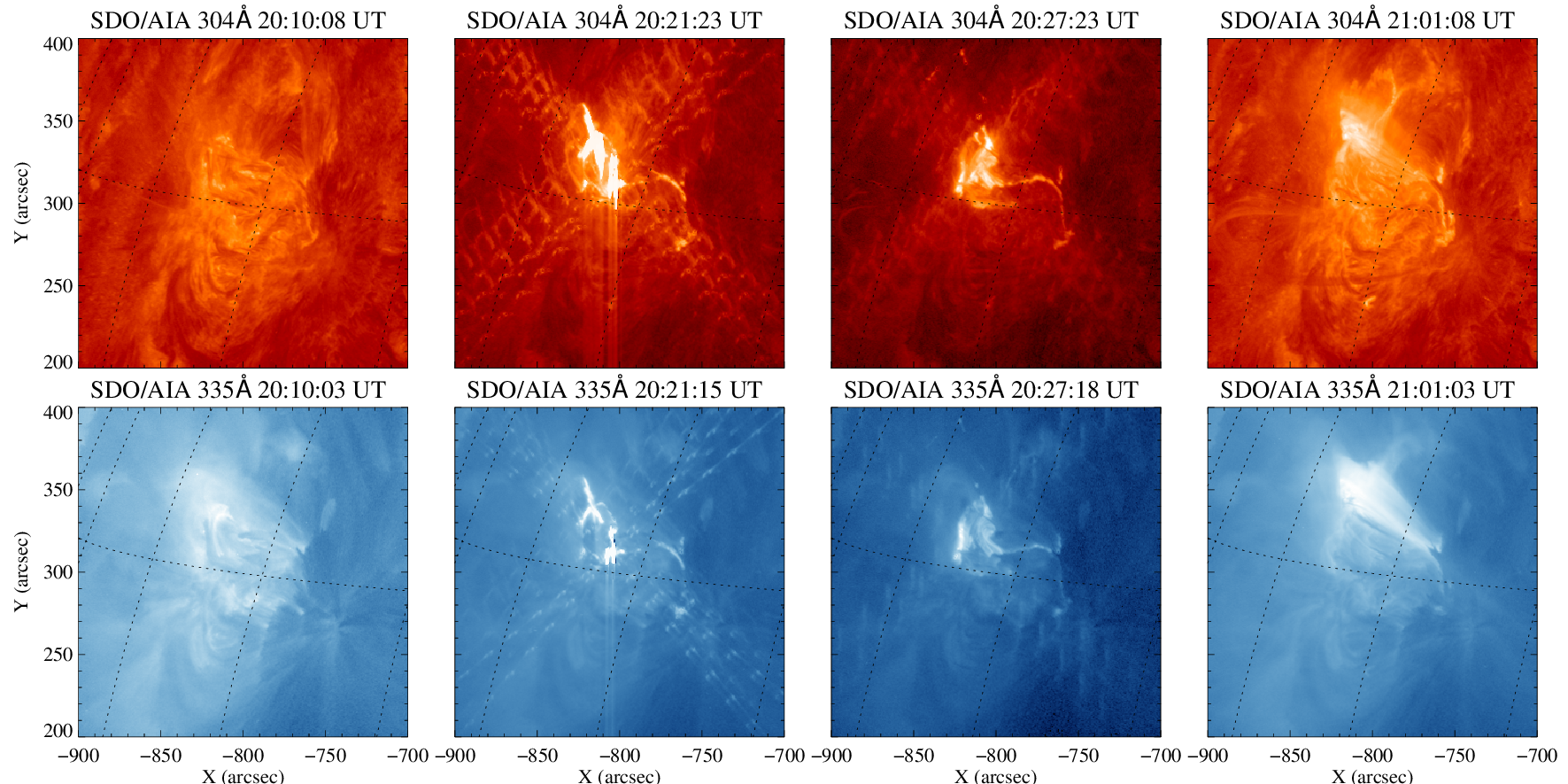}
        \centering
        \caption{SDO/AIA images of \ion{He}{2} 30.4 nm (top panels) and \ion{Fe}{16} 33.5 nm (bottom panels) emission for the pre-flare moment (20:10 UT), at the impulsive phase (20:21 UT), at the gradual phase (20:27 UT), and at the late phase of the $X2.9$ flare (21:01 UT) on 2011 November 3.}
    \label{SolarDisk}
    \end{figure}

Considering the above criteria, the $X2.9$ flare on 2011 November 3, which had a noticeable EUV late phase approximately 40 minutes after the main emission peak, was selected for analysis in this paper. Figure~\ref{SolarDisk} shows portions of the solar disk obtained by the AIA (Atmospheric Imaging Assembly; \citealp{Lemen_etal_2012}) instrument of the SDO satellite for the \ion{He}{2} 30.4 nm (top panels) and \ion{Fe}{16} 33.5 nm (bottom panels) lines, which characterize the impulsive and late flare phases respectively. The first column shows images of the active region before the flare (20:10 UT), the second column shows images during its impulsive phase (20:21 UT), when the main emission of cold chromospheric lines (such as \ion{He}{2} 30.4 nm, \ion{He}{1} 58.4 nm, and \ion{C}{3} 97.7 nm) was observed, the third column – during the gradual phase (20:27 UT), and the fourth column - during the EUV late phase (21:01 UT), when a significant increase in radiation was observed in warm coronal lines of the solar spectrum (\ion{Fe}{16} 33.5 nm and \ion{Fe}{15} 28.4 nm). As evident from bottom panels, the size of post-eruptive loops is indeed larger than the size of post-flare loops.

\begin{figure} [h]
        \noindent\includegraphics[width=490pt]{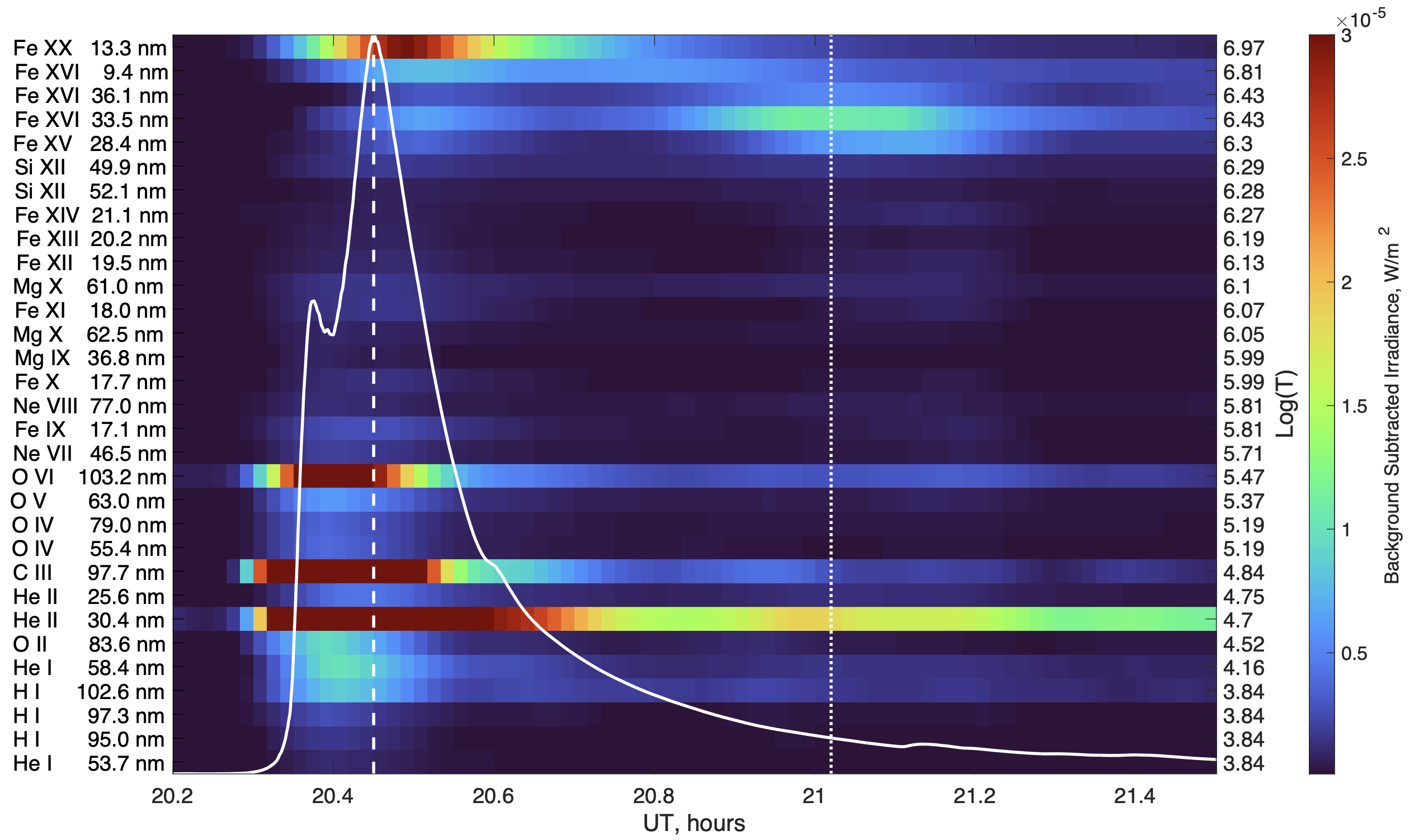}
        \caption{Relative irradiance in 31 lines measured by the SDO/EVE during the $X2.9$ flare on 2011 November 3. The emission lines are ordered by the logarithm of the characteristic temperatures which are shown on the right. Color indicates background subtracted irradiance values [$W/m^2$]. Solid white line indicates the normalised GOES $X$-ray flux (0.1-0.8 nm); vertical dashed line indicates the time moment of maximum $X$-ray flux; vertical dotted line indicates the time moment of maximum \ion{Fe}{16} 33.5 nm flux.}
    \label{IrradianceColor}
    \end{figure}

Figure~\ref{IrradianceColor} shows the emission variations relative to background values for 31 solar spectrum lines measured by the SDO/EVE during various phases of the $X2.9$ solar flare. According to data from the GOES satellite (Geostationary Operational Environment Satellite; \citealp{Machol_Viereck_2016}), the maximum $X$-ray flux in the range of 0.1–0.8 nm was reached at 20:27 UT; this moment is marked by the vertical dashed line. Solid white line indicates the normalised $X$-ray flux. In Figure~\ref{IrradianceColor}, we can see two peaks of warm coronal emissions (\ion{Fe}{16} 33.5 nm and \ion{Fe}{15} 28.4 nm). The first of these peaks ($\sim$ 20:29 UT) is the result of the plasma cooling after the maximum of the hot $X$-ray flux, so we can observe successive emissions of iron lines as the logarithm of temperature decreases from about 7 to 6.3 over a period of several minutes. The second emission peak is stronger and corresponds to the EUV late phase of the flare; the maximum emission flux of \ion{Fe}{16} 33.5 nm occurred at approximately 21:01 UT, this moment is shown by the vertical dotted line.

\begin{figure} [h]
        \noindent\includegraphics[width=260pt]{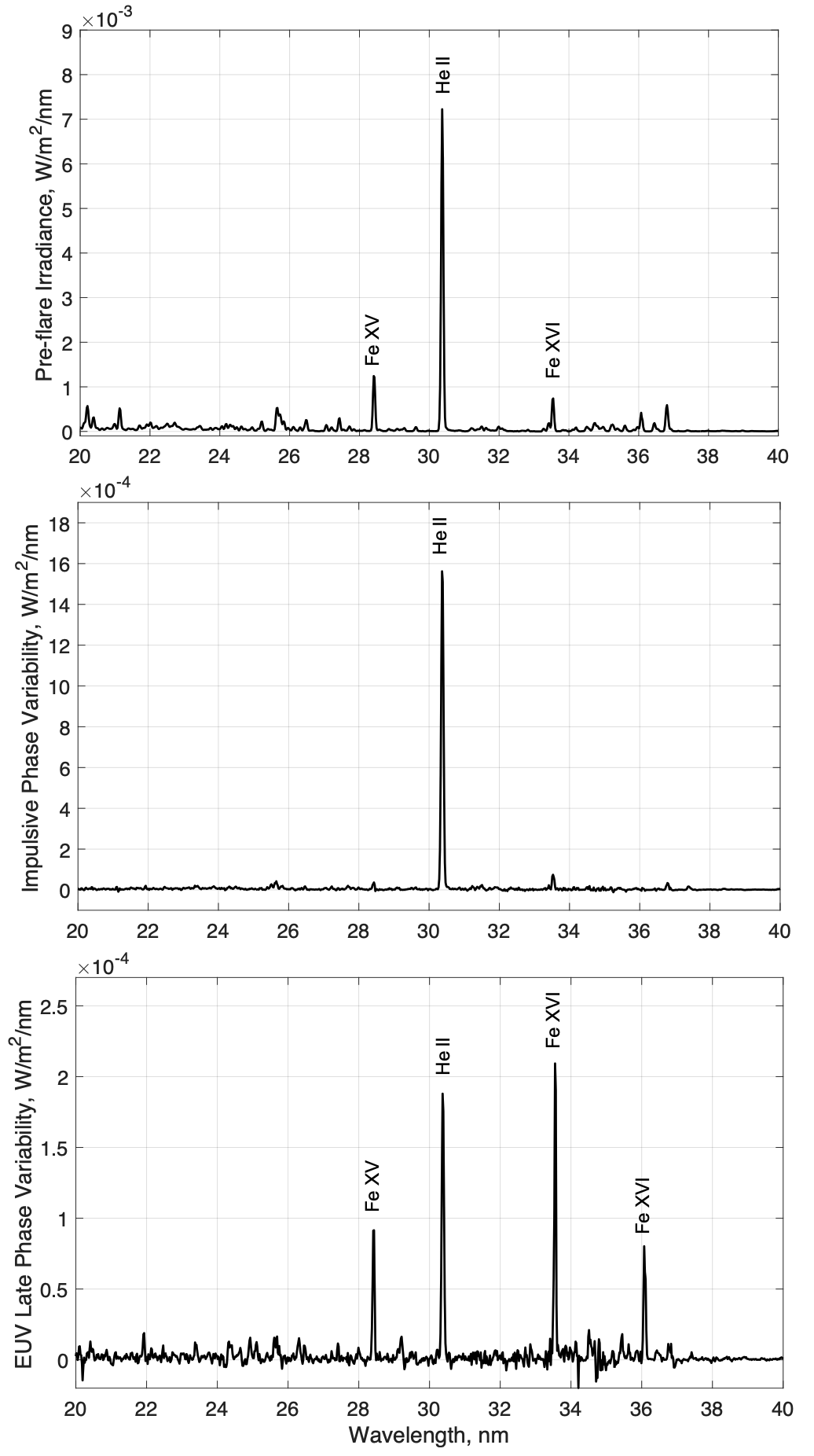}
        \centering
        \caption{Flare spectral variations from the SDO/EVE for the $X2.9$ flare on 2011 November 3. Top panel shows the pre-flare spectrum (20:18 UT). Middle panel shows the variability between the pre-flare irradiance (20:18 UT) and the impulsive phase (20:21 UT). Bottom panel shows the variability between the pre-flare irradiance (20:18 UT) and the EUV late phase (21:02 UT).}
    \label{Spectra}
    \end{figure}

Figure~\ref{Spectra} presents flare spectral variations from the SDO/EVE for the $X$2.9 flare on 2011 November 3. Top panel shows the pre-flare spectrum (20:18 UT), middle and bottom panels show the EUV spectral variations relative to the pre-flare values (20:18 UT) during the impulsive (20:21 UT) and late (21:02 UT) phases of this flare, respectively. As can be seen from figure, during the late phase, pronounced emission were observed in \ion{Fe}{15} 28.4 nm, \ion{Fe}{16} 33.5 nm, and \ion{Fe}{16} 36.1 nm lines in addition to commonly observed strong \ion{He}{2} 30.4 nm emission. While \ion{He}{2} emission comes from the post-flare loops, the warmer \ion{Fe}{15} and \ion{Fe}{16} emissions come from the post-eruptive loops as shown in right most panels of the Figure~\ref{SolarDisk}. Similar spectral profiles are presented in \citet{Woods_etal_2011, Liu_etal_2015}.

\subsection{Geoeffective lines of solar radiation}

It is known that the influence of various lines and continua of the solar spectrum on variations in the concentration of charged particles in the Earth’s ionosphere depends not only on the intensity of the emissions, but also on other geophysical factors. First, the energy of a particular photon must be large enough to overcome the ionization threshold of the neutral molecule. For example, the contribution of the strongest line in the solar spectrum, Lyman-$\alpha$ (121.6 nm), to the formation of the total electron content is practically not noticeable. This is because the energy of Lyman-$\alpha$ photons (10.2 eV) is sufficient to ionize only NO molecules (9.3 eV) in the Earth’s atmosphere, the concentration of which is many orders of magnitude lower than the concentration of the main neutral components (O, O$_2$, N$_2$). Therefore, the second factor in the geoeffectiveness of a solar spectrum line is the vertical profile of the concentration of the particle which is expected to be ionized. And thirdly, the ionization cross sections for specific particles and wavelengths play an important role in ionization processes. Modern theoretical models \citep{Solomon_Qian_2005,Watanabe_etal_2021,Nishimoto_2023} make it possible to calculate the altitude profile of the ionization rate under various heliogeophysical conditions, considering the above factors, and theoretically determine the most geoeffective solar radiation lines. Based on these estimates and variations in the spectrum of the flare in question ($X2.9$, 2011 November 3), six EUV lines were selected for analysis, which are most likely responsible for the increase in the electron content in the ionospheric $E$ and $F$ regions during various phases of the flare. The chosen EUV lines are presented in Table~\ref{Wl}. The wavelength ranges listed in the table correspond to the measurement range of the EVE instrument for a given wavelength. Together with a set of EUV emissions, the $X$-ray range of 0.1–0.8 nm (GOES), which is responsible for ionization of the ionospheric $D$ region during flares, was considered. Despite the relatively low electron concentration in the $D$ region, it can play an important role in the $TEC$ increment during a flare. \citet{Bekker_Ryakhovsky_2024} showed that the contribution of the 50–90 km altitude range to the $TEC$ increment during $X$-class flares can be up to 23\%, depending on the fraction of $X$-ray emission in the spectral composition of the flare.

\begin{table}[h]
    \caption{The Characteristics of the Geoeffective Lines of Solar Radiation Used in This Study}
    \centering
        \begin{tabular}{l c c c}
        \hline
        \hline
        Ion & Wavelength centre [nm] & Wavelength Range [nm] & Log(T, [K]) \\
        \hline
            \ion{Fe}{10} &	13.28 & 13.23–13.32 & 6.97  \\
            \ion{Fe}{15} &	28.41 & 28.30–28.50 & 6.30  \\
            \ion{He}{2} &	30.37 & 30.25–30.50 & 4.70  \\
            \ion{Fe}{16} & 33.54 & 33.49–33.61 & 6.43  \\
            \ion{He}{1} & 58.43 & 58.39–58.51 & 4.16 \\
            \ion{C}{3} & 97.70 & 97.65–97.77 & 4.84  \\
        \end{tabular}
        \label{Wl}
\end{table}

\begin{figure} [h]
        \noindent\includegraphics[width=480pt]{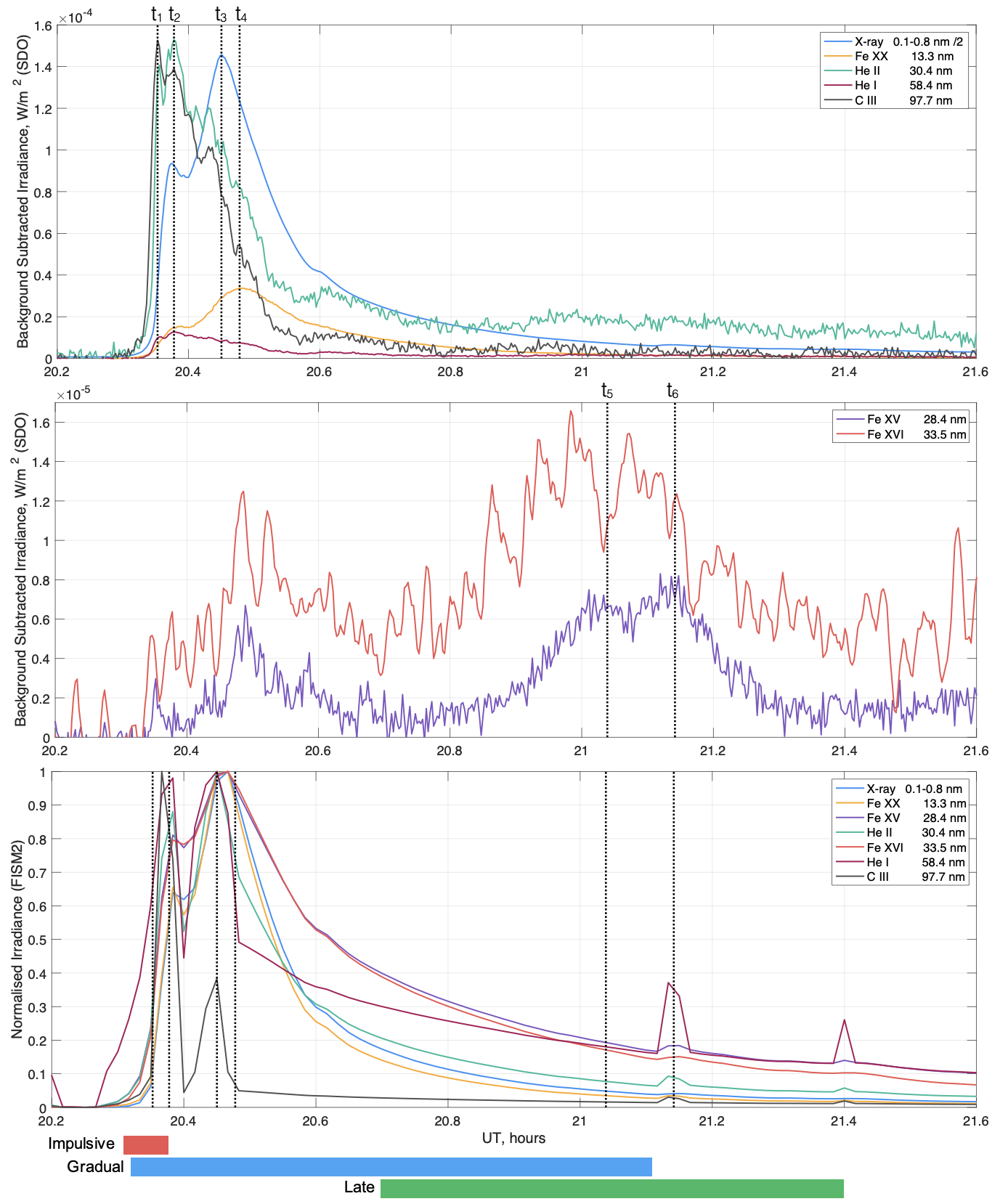}
        \centering
        \caption{Top panel: Background subtracted solar flare lightcurves in hot coronal and chromospheric emissions from GOES and SDO/EVE during the $X2.9$ flare on 2011 November 3. Middle panel: Background subtracted solar flare lightcurves in warm coronal emission lines from SDO/EVE during the $X2.9$ flare on 2011 November 3. Bottom panel: Normalised values of the same emission lines as predicted by FISM2 model. Horizontal colorbars indicate the time ranges for the impulsive (red), gradual (blue), and late (green) phases of the flare. Vertical dotted lines correspond to the times of the main emission peaks that are listed in the Table 2.}
        \label{Irradiance}
    \end{figure}

Figure~\ref{Irradiance} shows the time variations of radiation fluxes in selected lines of the solar spectrum listed in Table~\ref{Wl}. The top panel shows the background subtracted irradiance values measured by the SDO and GOES satellites (for convenience, the absolute values of $X$-ray flux are divided by 2) for the hot coronal ($\gtrsim 10^7$ K) and chromospheric emissions ($10^4$$-$$10^5$ K). The middle panel shows the background subtracted irradiance values measured by the SDO for the warm coronal emissions ($10^6$$-$$10^7$ K). The bottom panel shows predicted normalized radiation fluxes obtained from the empirical FISM2 model (Flare Irradiance Spectral Model; \citealp{Chamberlin_etal_2020}).

The top and middle panels of Figure~\ref{Irradiance} clearly reveal all three phases of the solar flare:

\begin{itemize}
  \item[\sbt] impulsive phase, when the main peaks of EUV and $X$-ray radiation and the maximum growth rate of soft $X$-ray are observed; the main source of this radiation are the footpoints located in the solar chromosphere (red horizontal colorbar);
  \item[\sbt] a gradual phase, when soft $X$-ray/EUV radiation reach their maxima and then decay; due to the heating  and cooling of the rising chromospheric plasma, loop arcades arise at this stage (blue horizontal colorbar);
  \item[\sbt] late phase, which is characterized by a significant increase in radiation in the warm coronal lines of the solar spectrum, such as \ion{Fe}{16} 33.5 nm ($T$ = $10^{6.43}$ K) and \ion{Fe}{15} 28.4 nm ($T$ = $10^{6.30}$ K) (green horizontal colorbar).
\end{itemize}

The results of the FISM2 model are commonly used as input parameters of plasma-chemical models to calculate the response of ionospheric components to variations in solar radiation in different spectral ranges \citep{Pettit_etal_2018, Qian_etal_2019}. At the same time, it was previously shown that the FISM2 model can underestimate EUV fluxes during flares \citep{Greatorex_etal_2023}. And, as can be seen from the bottom panel of Figure~\ref{Irradiance}, in addition to the inaccuracies observed during the impulsive and gradual phases, FISM2 fails to predict EUV late phase emissions, which is much longer and consequently its energy can be similar to the main phase energy \citep{Woods_etal_2011,Liu_etal_2024}. Despite the importance and wide applicability of this model, these results should be taken into account when using FISM2 to predict the dynamics of ionospheric layers under the influence of solar radiation.

For a further analysis of the total electron content response to variations in solar radiation, six moments of time that correspond to key emissions at different phases of the flare under consideration were selected. These moments are marked in Figure~\ref{Irradiance} with vertical dotted lines and correspond to the times and spectral lines listed in Table~\ref{Time}.

\begin{table}[h]
 \caption{Moments of Time of the Main Emission Peaks During Different Phases of $X2.9$ Solar Flare on 2011 November 3}
    \centering
        {\begin{tabular}{c c c c c c}
        \hline
        \hline
        $t_1$ & $t_2$ & $t_3$ & $t_4$ & $t_5$ & $t_6$ \\
        \hline
        \multicolumn{2}{c}{\ion{He}{2}, \ion{He}{1}, \ion{C}{3}} &	$X$-ray & \ion{Fe}{10} &  \multicolumn{2}{c}{\ion{Fe}{15}} \\
            \hline
            20:21 UT &	20:23 UT & 20:27 UT & 20:29 UT & 21:02 UT & 21:08 UT \\
        \end{tabular}}
        \label{Time}
\end{table}

The first two moments correspond to the main maxima of cold chromospheric lines (\ion{He}{2}, \ion{He}{1}, \ion{C}{3}), the 3$^{rd}$ and 4$^{th}$ – to hot coronal lines ($X$-ray, \ion{Fe}{10}), the 5$^{th}$ and 6$^{th}$ – to the maxima of the warm coronal line, \ion{Fe}{15}.

\section{Ionospheric data}
\subsection{Data of Global Navigation Satellite Systems}

\begin{figure} [h]
        \noindent\includegraphics[width=245pt]{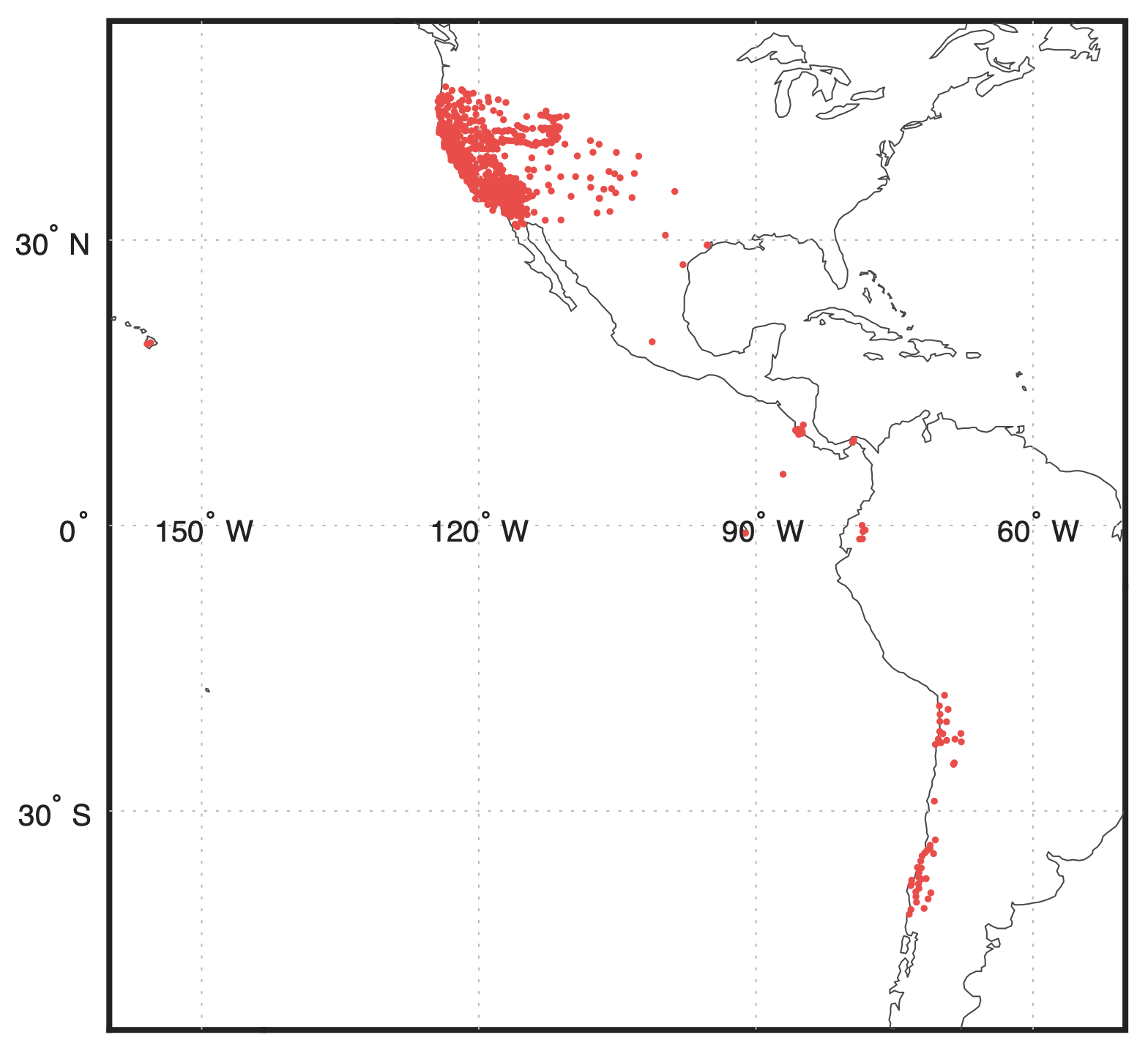}
        \centering
        \caption{Location of the GNSS stations of the SOPAC network used in this work.}
        \label{Map}
    \end{figure}

As discussed in the Introduction section, sudden increases in EUV and $X$-ray fluxes lead to perturbation, absorption, and delay of signals of various frequencies, which let us use data from GNSS to estimate the dynamics of the total electron content during various phases of a solar flare. To assess the impact of various solar flare emissions on the ionospheric dynamics, we used data accumulated from the SOPAC (Scripps Orbit and Permanent Array Center) network stations. Variations in solar radiation led to increased ionization of the illuminated part of the Earth’s ionosphere; therefore, we selected data only from those stations where the solar zenith angle (SZA) during the flare was less than 60°. We also limited the latitude range from –45° to +45° to exclude effects associated with particle precipitation and other processes occurring in polar ionosphere. As a result, data from 956 stations were selected, their locations are shown in red in Figure~\ref{Map}. On average, 4–5 satellites were simultaneously observed at each station, which led to the analysis of several thousand satellite flybys.

\subsection{TEC data processing}

\begin{figure} [h]
        \noindent\includegraphics[width=450pt]{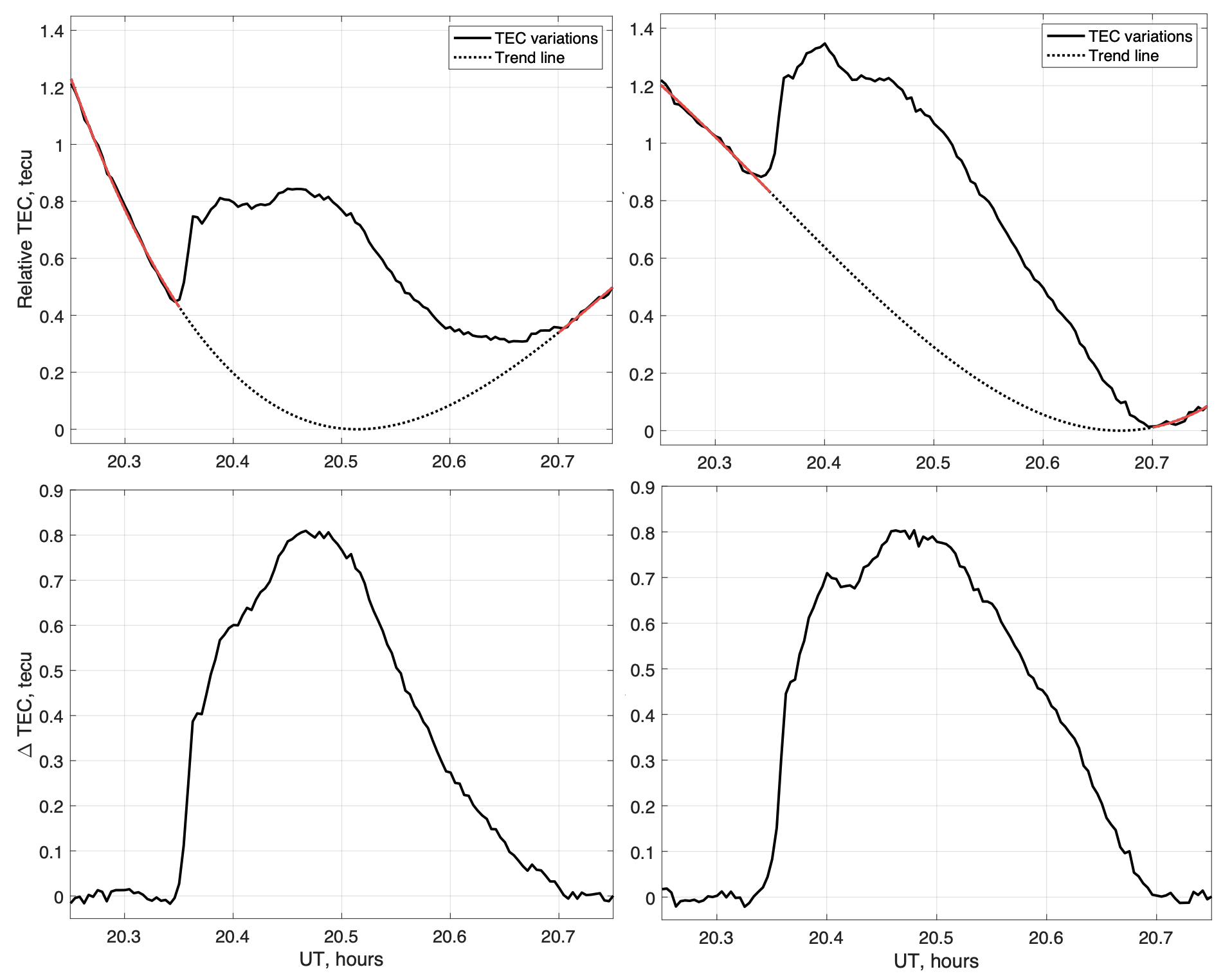}
        \centering
        \caption{Top panels: The relative $TEC$ values (solid lines) and the subtracted trends (dotted lines) obtained at “ana1” (34.0°N 119.4°W; left) and “arm1” (35.2°N 118.9°W; right) GPS stations during the main phase of the $X2.9$ flare on 2011 November 3. Red color indicates the time ranges used for parabolic trends construction. Bottom panels: Corresponding $\Delta TEC$ dynamics during the main phase of the flare.}
        \label{SatImp}
    \end{figure}

To numerically estimate the increment in $\Delta TEC$ during different phases of the $X2.9$ solar flare, parabolic trends associated with satellite trajectory and differential code biases were subtracted from the measured total electron content values. The black solid lines in the top panels of Figure~\ref{SatImp} show the measured relative $TEC$ variations during the main phase of the flare. These data were obtained from the GPS stations “ana1” (34.0°N 119.4°W; left) and “arm1” (35.2°N 118.9°W; right); dotted lines here show background parabolic trends that do not contain any information about changes in solar radiation; red color indicates the time ranges used to extrapolate the trend lines. These time ranges were selected in such a way that the vertical $\Delta TEC$ values obtained at one station approximately coincided for all satellites of the constellation \citep{Bekker_Ryakhovsky_2024, Taurenis_etal_2022}. In the bottom panels of Figure~\ref{SatImp}, we can see $\Delta TEC$ value obtained by subtracting the dotted curves from the solid curves.

\begin{figure} [h]
        \noindent\includegraphics[width=450pt]{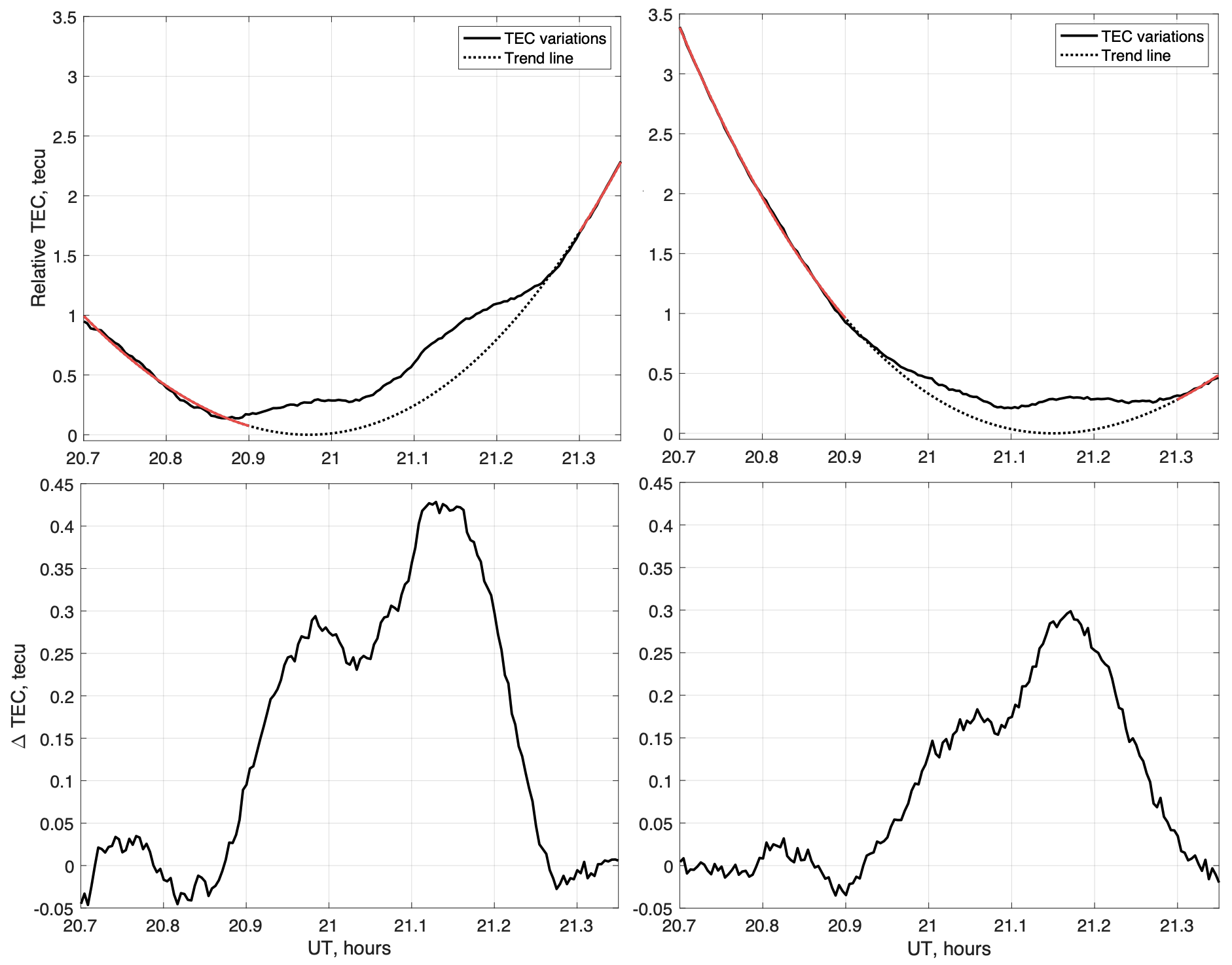}
        \centering
        \caption{Top panels: The relative $TEC$ values (solid lines) and the subtracted trends (dotted lines) obtained at “bbdm” (34.6°N 120.0°W; left) and “cast” (39.2°N 110.7°W; right) GPS stations during the EUV late phase of the $X2.9$ flare on 2011 November 3. Red color indicates the time ranges used for parabolic trends construction. Bottom panels: Corresponding $\Delta TEC$ dynamics during the EUV late phase of the flare.}
        \label{SatLate}
    \end{figure}

Figure~\ref{SatLate} presents the similarly estimated $\Delta TEC$ increment during the late phase of the flare at the “bbdm” (34.6°N 120.0°W; left) and “cast” (39.2°N 110.7°W; right) GPS stations. Due to the slower rise in ionization during the late phase, the detrending process was more difficult. Therefore, to minimize detrending errors, calculations were carried out only for satellites whose locations during the flare were approximately above the ground receivers.

As expected, due to the non-uniform spatial distribution of GPS stations (most of them are on the west coast of the US), the obtained $\Delta TEC$ variations did not demonstrate a latitude-longitudinal dependence or a dependence on the solar zenith angle either during the main phase, or during the late phase.

\section{Results and discussion}

\begin{figure} [h]
        \noindent\includegraphics[width=480pt]{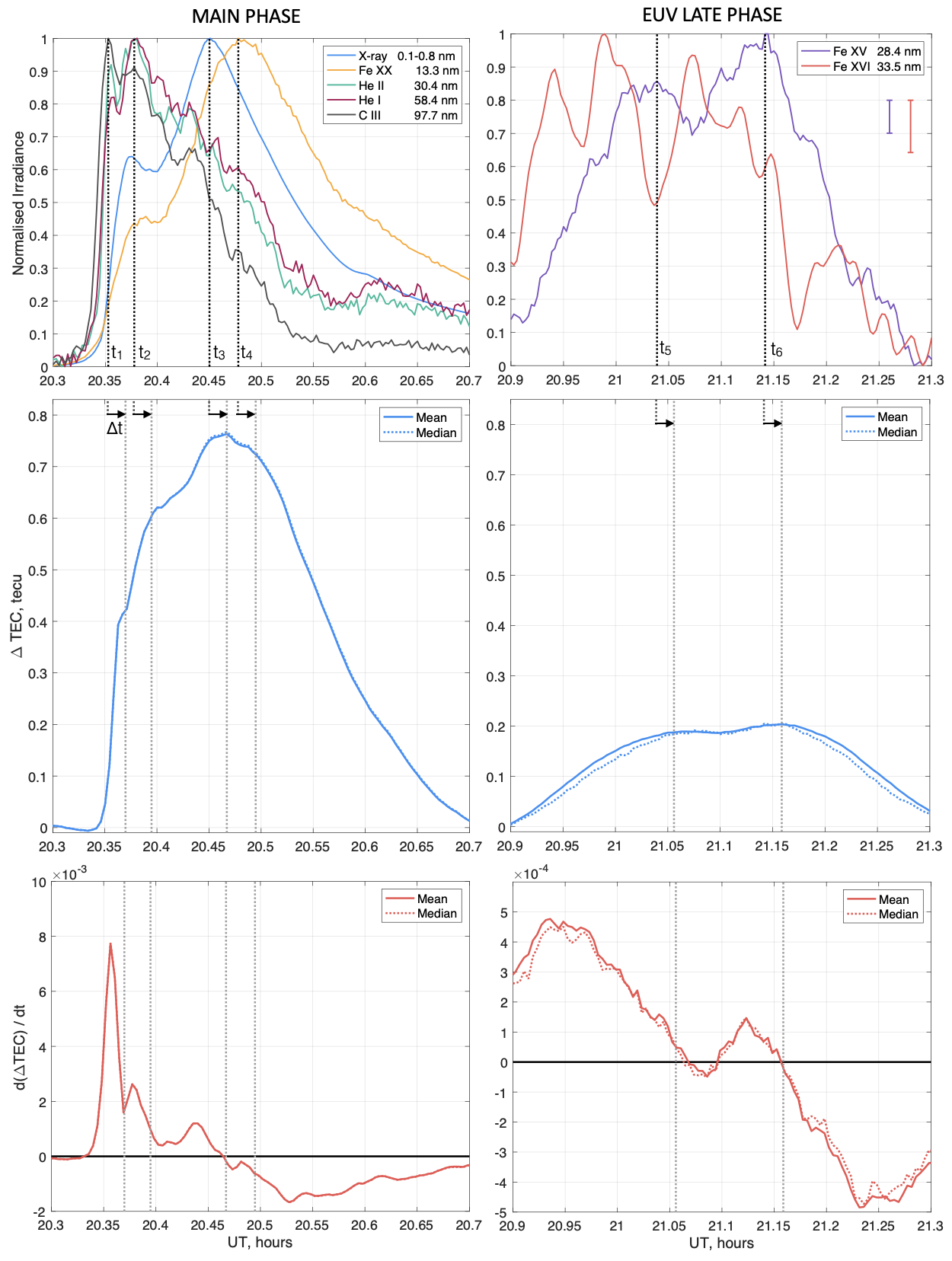}
        \centering
        \caption{Top panels: Solar flare lightcurves in $X$-rays and EUV emission lines during the main (left) and late (right) phases of the $X2.9$ flare on 2011 November 3. Middle panels: The corresponding $TEC$ response obtained by averaging data from 956 GPS stations. Bottom panels: The corresponding time derivative, d($\Delta TEC$)/dt.}
        \label{dTEC}
    \end{figure}

Following the methodology described in Section 3.2, $TEC$ measurements obtained at 956 selected GNSS stations were detrended to calculate clear $\Delta TEC$ increment. The top panels of Figure~\ref{dTEC} show the main and the most geoeffective emissions (according to theoretical estimates) during the main (left) and late (right) phases of the solar flare under consideration. The top right panel presents smoothed flux data that may be responsible for the $TEC$ increment during the EUV late phase. Size of the sliding window (1 min) was chosen in such a way that the resulting curve correctly presents the dynamics of the experimental measurements. Errorbars for the EUV late phase emissions are shown on the right part of this panel below the legend. The middle and bottom panels present the resulting averaged dynamics of the $\Delta TEC$ and the time derivative d($\Delta TEC$)/dt for the same time ranges. As evident from Figure~\ref{dTEC}, the average response of the ionosphere to the EUV late phase of the flare was almost 30\% of the response to the significantly more powerful impulsive phase. So, we can draw at least 2 conclusions:
\begin{itemize}
  \item[\sbt] late warm coronal emissions have quite high geoeffectiveness (despite the fact that their absolute flux values are an order of magnitude lower than the flux values of cold chromospheric lines);
  \item[\sbt] the previously ignored EUV late phase of a solar flare should be considered when modeling and predicting the $Ne$ response to variations in solar radiation, since it also causes a noticeable increase in $TEC$.
\end{itemize}

In addition, the time profile of the $\Delta TEC$ curve has inflections and maxima, which are most conveniently tracked by the decreasing segments of its derivative (bottom panels of Figure~\ref{dTEC}). For example, during the impulsive and gradual phases, $\Delta TEC$ has four local peaks corresponding to the main EUV and $X$-ray emissions discussed in Section 2.2 ($t_1$, $t_2$, $t_3$, $t_4$). The reaction of the electron concentration to an increase in radiation fluxes has a time delay $\Delta t$, which is approximately 1 minute. This time delay is due to the ionization-recombination balance at altitudes in the ionospheric $F$ region, where the bulk of the total electron content is contained. As can be seen, $Ne$ increases until the main emission of $X$-rays (0.1–0.8 nm), and it decreases after the moment $t_3+\Delta t$, since recombination processes begin to dominate over ionization. The energy of subsequent \ion{Fe}{10} 13.3 nm emission is sufficient only to slow down the falling rate of the electron concentration, but not to fundamentally change the shape of the $TEC$ curve.

During the EUV late phase of the flare, two clear $\Delta TEC$ maxima were detected, the second of which is stronger. Such $Ne$ response, both in shape and time, corresponds to the \ion{Fe}{15} 28.4 nm line (purple curve). Thus, the increase in the total electron content during the late phase is most likely associated with the \ion{Fe}{15} 28.4 nm, rather than the main characteristic of the EUV late phase of a flare – \ion{Fe}{16} 33.5 nm \citep{Woods_etal_2011}.

\begin{figure} [h]
        \noindent\includegraphics[width=510pt]{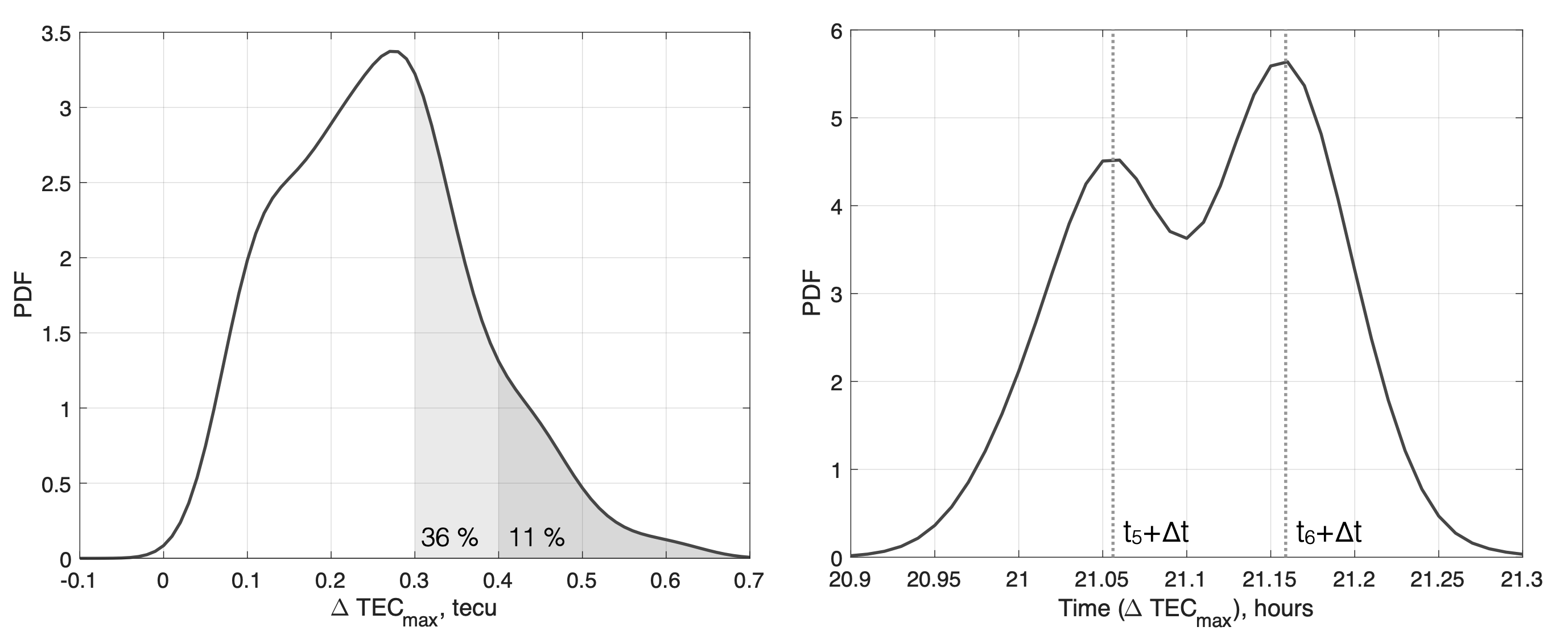}
        \caption{Probability density functions of $\Delta TEC_{max}$ (left) and of moment of time when the $\Delta TEC$ reached its maximum (right) during the EUV late phase of the $X2.9$ flare on 2011 November 3.}
        \label{PDF}
    \end{figure}

As we can see, the average response to the EUV late phase of the $X2.9$ flare was about 0.2 tecu, but at some stations it was significantly higher. For example, it exceeded 0.3 tecu in 36\% of cases, and 0.4 tecu in 11\% of cases. This can be seen in the left panel of Figure~\ref{PDF}, which presents the probability density function of the maximum increment, $\Delta TEC_{max}$, during the late phase of the flare. The right panel shows the probability density function of the moments of time when this maximum value was reached. In this panel, the correlation with the \ion{Fe}{15} 28.4 nm emission is even more clear: measurements at most stations showed that the second peak of the emission was indeed stronger. This fact additionally confirms the hypothesis of a considerable contribution of this particular line to the ionization of the ionosphere during the EUV late phase of the flare.

The demonstrated total electron content response to the late phase of this solar flare is significantly lower than the value obtained by \citet{Liu_etal_2024} for a flare of approximately the same magnitude. Such difference may be due to at least two reasons. The result presented here was calculated using data from stations with solar zenith angles from 0° to 60°, and not for the subsolar point. Additionally, when assessing the ionospheric response to the EUV late phase, our method enabled us to avoid effects associated with increased background ionization after the impulsive phase and demonstrated a $TEC$ increase caused exclusively by the late emission of warm coronal lines, therefore the estimated value is much smaller.

\section{Conclusions}

The study of the influence of solar flares on the processes occurring in the Earth's ionosphere remains an important and relevant problem today. It has been repeatedly shown that the illuminated part of the Earth's ionosphere is very sensitive to variations in solar radiation fluxes, which can cause failures in communication and navigation systems. Further study of the complex dynamics of ionospheric layers under the influence of increased ionization is necessary to improve the accuracy of modeling and forecasting such events.

There are different approaches to the study of solar-terrestrial connections: purely theoretical \citep{Meza_etal_2006,Bekker_etal_2021,Bekker_etal_2022,Yan_etal_2022}, hybrid \citep{Hazarika_etal_2016,Qian_etal_2019,Nishimoto_2023,Ryakhovsky_etal_2023}, or empirical \citep{Barta_etal_2019,Gavrilov_etal_2019a,Habarulema_etal_2022,Kolarski_etal_2022,Nina_2022,Buzas_etal_2023}. Observations of synchronous variations in solar radiation and ionospheric parameters is a rather promising research direction, since it allows us to experimentally answer some questions about the geoeffectiveness of certain emissions, the delay in the ionospheric reaction and, of course, to numerically estimate the magnitude of the electron concentration response to disturbances of various powers and spectra.

In this paper, we analyzed synchronous variations in solar radiation and total electron content during the $X$-class solar flare that occurred on 2011 November 3. In addition to the powerful impulsive phase, this event is characterized by a noticeable EUV late phase, during which repeated and stronger emissions of warm coronal radiation (\ion{Fe}{15}, \ion{Fe}{16}) occurred. The SDO/EVE instrument measured the emission in 39 EUV lines at all phases of the flare with a 10-second cadence. The availability of these data together with measurements from the GOES satellite enabled to compare in detail the dynamics of the measured total electron content with successive emissions of cold chromospheric lines (\ion{He}{2} 30.4 nm, \ion{C}{3} 97.7 nm, and \ion{He}{1} 58.4 nm), hot $X$-rays (0.1–0.8 mn) and coronal emissions (\ion{Fe}{10} 13.3 nm), and warm coronal emissions of the EUV late flare phase (\ion{Fe}{15} 28.4 nm and \ion{Fe}{16} 33.5 nm).

The $TEC$ dynamics in this work was assessed using data from 956 GNSS stations. Available measurements of GPS signals allowed us to calculate $TEC$ variations with 15-second cadence during different phases of a solar flare. As a result of analysis of the obtained averaged $TEC$ variations, we identified several maxima in the electron density increment, which correspond in time to the key emissions of the solar flare. It turned out that the production of charged particles in the ionosphere continues after a sharp impulsive increase in EUV radiation and reaches a maximum only after the peak of hard $X$-ray radiation, which ionizes the lower part of the ionosphere ($D$ region and partially $E$ region). As $X$-ray radiation declines, recombination processes begin to prevail over ionization processes; subsequent emission affect only the rate of $Ne$ recession, but do not fundamentally change the shape of the total electron content curve.

An important part of this study is the numerical assessment of the $Ne$ response to the EUV late phase of the flare. To date, it has not attracted much attention of the scientific community, despite the fact that a second powerful peak of warm emissions with longer coronal loops than during the impulsive phase occurs in almost half of $X$-class flares \citep{Woods_Thomas_2014}. It was found that the ionospheric response to the relatively weak emissions of the EUV late phase of the $X2.9$ solar flare amounted to almost a third (more than 0.2 tecu) of the $TEC$ increase during the impulsive phase. Analysis of the time profile of the total electron content showed that this is most likely caused by the high geoeffectiveness of the \ion{Fe}{15} 28.4 nm emission, which mostly ionizes O and N$_2$ molecules in the ionospheric $F$ region \citep{Watanabe_etal_2021,Nishimoto_2023}.  The energy of the EUV late phase of more powerful event than those considered in this work can be much higher, so the obtained result indicates a serious need to consider the late emission of warm coronal lines when modeling and forecasting the ionospheric response to variations in solar radiation. Unfortunately, the most well-known and widely used model of the solar spectrum, FISM2, fails to predict this effect.

The results presented in this work demonstrated the great promise of analysis of synchronous variations in solar spectrum radiation and experimental measurements of ionospheric parameters for the study of solar-terrestrial connections. Therefore, it is planned to continue the present study using the techniques proposed here on a larger number of events of varying power and spectral composition.

\section{Open Research}

No new data were generated as part of this research. Data from the SOPAC network are available at \url{http://sopac-old.ucsd.edu/}. The FISM2 model data can be obtained under \url{https://lasp.colorado.edu/eve/data_access/eve_data/fism/}. The GOES and SDO data used in this study are available at \url{https://www.ncei.noaa.gov/data/goes-space-environment-monitor/} and \url{https://lasp.colorado.edu/eve/data_access/}, respectively.

\section{Acknowledgments}

S.Z.B. and R.O.M. would like to thank the European Office of Aerospace Research and Development (FA8655-22-1-7044-P00001) for supporting this research. R.O.M. would also like to acknowledge support from STFC New Applicant grant ST/W001144/1. I.A.R. would like to acknowledge support from Ministry of Science and Higher Education of Russian Federation (122032900175-6).

\bibliography{sample631}{}
\bibliographystyle{aasjournal}

\end{document}